\documentclass[sigconf]{aamas}  % do not change this line!

\usepackage{booktabs}
\usepackage{amsmath}
\usepackage{algorithm}
\usepackage{url}
\usepackage{listing}
\usepackage[noend]{algpseudocode}

\usepackage{flushend}
\setcopyright{rightsretained}  % do not change this line!
\acmDOI{}  % do not change this line!
\acmISBN{}  % do not change this line!
\acmConference[~]{This is the full version of the extended abstract published at AAMAS}{2020}{\url{https://dl.acm.org/doi/abs/10.5555/3398761.3398981}}  % do not change this line!
\acmYear{2020}  % do not change this line!
\copyrightyear{2020}  % do not change this line!
\acmPrice{}  % do not change this line!

\newcommand{\galt}{\ensuremath{\mid}~}
\newcommand{\FORGET}[1]{}

\newcommand{\todo}[1]{}
\newcommand{\labelsec}[1]{\label{sec:#1}}

\begin{document}

%\title{Encapsulating Reactive Behaviour in Goal-based Plans \\ in BDI Agents Programming}
\title[~]{Encapsulating Reactive Behaviour in Goal-based Plans for Programming BDI Agents}
%%RHB: Removed \\ otherwise it breaks a line in the heading too

%\titlenote{Produces the permission block, and copyright information}

% AAMAS: as appropriate, uncomment one subtitle line; check the CFP
%\subtitle{Extended Abstract}
%\subtitle{Blue Sky Ideas Track}
%\subtitle{JAAMAS Track}
%\subtitle{Demonstration}
%\subtitle{Doctoral Consortium}

% AAMAS: submissions are anonymous for most tracks
%\author{Paper \#1075}  % put your paper number here!

%% example of author block for camera ready version of accepted papers: don't use for anonymous submissions
%
\author{Rafael H. Bordini}
%\authornote{The secretary disavows any knowledge of this author's actions.}
\affiliation{%
  \institution{School of Technology, PUCRS}
  \streetaddress{Av. Ipiranga, 6681}
  \city{Porto Alegre, RS} 
  % \postcode{90619-900}
  \state{Brazil} 
}
\email{rafael.bordini@pucrs.br}

\author{Rem Collier}
\affiliation{%
  \institution{University College of Dublin}
  \city{Dublin}
  \country{Ireland}
}
\email{rem.collier@ucd.ie}
\author{Jomi F. H\"ubner}
% \authornote{This author is the
%  one who did all the really hard work.}
\affiliation{%
  \institution{DAS, Federal University of Santa Catarina}
  \streetaddress{PO Box 476}
  \city{Florianópolis, SC} 
  %\postcode{88040-900}
  \country{Brasil}
}
\email{jomi.hubner@ufsc.br}
\author{Alessandro Ricci}
%\orcid{1234-5678-9012}
\affiliation{%
  \institution{DISI, University of Bologna}
  \streetaddress{Via Sacchi, 3}
  \city{Cesena} 
  \state{Italy} 
  % \postcode{47521}
}
\email{a.ricci@unibo.it}
\renewcommand{\shortauthors}{~}

\begin{abstract}
Reactive behaviour in BDI-based models and architectures adopted in agent programming is typically specified in terms of reactive plans not bound to any specific goal. In this paper, we present and discuss an extension of the plan model used in BDI programming languages in which goal-based plans encapsulate both proactive and reactive behaviour. This brings important benefits both to the practice of agent programming and in supporting agent reasoning at runtime. The paper first introduces the model formally, abstracting away from particular programming languages. The approach is then evaluated through concrete implementations based on two existing agent programming platforms, namely Jason and ASTRA. The paper reports on experiments showing that the approach provides for elegant programming and can also be efficient.
 \end{abstract}

  % put your semicolon-separated keywords here!

\maketitle

\section{Introduction}
\labelsec{introduction}

The BDI architecture and model has been either the basis or at least an important influence in many agent programming languages. This has led to many concrete computational models and platform implementations, even if the basic programming language is a traditional one, for example PRS~\cite{Ingrand:1992:ARR:629535.629890}, dMARS~\cite{D'Inverno:2004:DAS:974009.974022}, JAM~\cite{Huber:1999:JBM:301136.301202}, JACK~\cite{DBLP:books/sp/map2005/Winikoff05}, and SPARK~\cite{Morley:2004:SAF:1018410.1018821}, to name just a few. It also leads to the creation of various programming languages and variants of those languages, whether abstract (and used mostly for formalisation of ideas related to BDI) or more practical, for example AgentSpeak(L)~\cite{Rao96}, CAN~\cite{Sardina2011}, Jason~\cite{jason06}, ASTRA~\cite{DBLP:conf/prima/CollierRL15}, and Gwendolen~\cite{dennis08:_gwend}, again to name just a few.

Needless to say, plans have a key role in programming BDI agents, regardless of the particular approach. Plans define the agent know-how, they express a course of action that can be used to bring about a state-of-affairs, particularly those that are desirable to the agent, hence plans have a strong relation to goal-orientation. A plan has either a goal to achieve or to maintain and so they form an essential part of agents' behaviour.

However, the plan model adopted in all concrete BDI-based computational platforms, from the very early days, have pitfalls that long-term experience has helped surface. One of them is a lack of structure with the plan library, and in particular lack of encapsulation. A plan library is usually just a collection of plans to achieve individual (top-level) goals and their subgoals. However, the intentional context in which a subgoal needs to be pursued is relevant, that is, in many cases the agent has few top-level goals, and subgoals might need to be pursed differently depending on which top-level goal has originated the subgoal. Another typical problem is that, because agents need to be reactive as well as proactive, it has been tempting to just use a plan programming construct to specify reactive behaviour, but again these may vary significantly depending on the context of the overall top-level goals that the agent is currently trying to achieve.

This paper puts forward an approach to address these problems. It preserves goal-orientation while being cleanly event-driven. Most importantly, it increases encapsulation in defining the strategies for plans to achieve/maintain goals. This not only has a clear impact both at the design and programming activities, but also for reasoning about gals and intentions at runtime. In particular, we conjecture it may have an impact in the intention progression problem~\cite{Logan:2017:PIP:3091125.3091234}.

We first present our extended plan model formally and in a language-independent way. Then, we present implementations of our approach extending two different existing BDI platforms, namely Jason and ASTRA. We show experimental results to assess how the extended platforms compare to the original ones. We then discuss how this approach addresses the pitfalls of traditional BDI plan models.

\section{Background and Motivation}
\labelsec{background}

The Belief-Desire-Intention (BDI) architecture is one of the major models adopted in academia and industry for developing intelligent agents and in particular practical reasoning agents.
One of the first concrete architectures based on BDI was PRS~\cite{Ingrand:1992:ARR:629535.629890}, implemented then by the dMARS system~\cite{D'Inverno:2004:DAS:974009.974022} and applied in several significant multi-agent applications.
A whole range of practical development efforts relating to BDI systems have been undertaken, either as refinements of PRS and dMARS or more loosely based on broader BDI principles, including PRS-Lite, JAM, JACK, Jadex.
Besides systems, several agent programming languages %in Agent-Oriented Programming
have been developed based on BDI (surveys can be found here~\cite{DBLP:books/sp/map05,DBLP:books/sp/FDDB2009}).
Among them, we mention here AgentSpeak(L)~\cite{Rao96} ---  an abstract language based on an abstraction of the PRS architecture, stripped down to its bare essentials, later extended and implemented by concrete Agent Programming platforms such as Jason and ASTRA --- and influencing many languages including the more recent CANPlan~\cite{Sardina:2011:BAP:1969591.1969594}, which is equipped with a formal specification.

In spite of specific differences, all BDI computational models are based on the same conceptual model for plans and intentions.
A \emph{plan} represents how to bring about a state of affairs.  
\emph{Intentions} represent a specific state of affairs %-- desires. \emph{goals} --
that the agent is committed to achieving and the activity used for that purpose.
A plan then is a recipe specifying the course of action that may be undertaken by an agent in order to achieve such states of affairs.
Plans are collected in agent's plan library, representing its \emph{procedural knowledge} or \emph{know-how}.

%
% THE COMPUT MODEL
%
The issues that we focus on this paper are about how this conceptual model is then reified in concrete computational and programming models.
Each plan is typically defined by some key components, including~\cite{D'Inverno:2004:DAS:974009.974022}:
\begin{itemize}
\item a \emph{trigger} or \emph{invocation} condition, which specifies the circumstances under which the plan should be considered relevant, usually specified in terms of \emph{events};
\item  a \emph{context}, or \emph{pre-condition}, specifying the circumstances under which the execution of the plan may commence and the plan considered applicable;
\item a \emph{body}, defining a potentially quite complex course of actions which may consist of both goals (or subgoals) and primitive actions.
\end{itemize}
This model is pervasively used, from the original dMARS specification up to the more recent CANPlan~\cite{Sardina:2011:BAP:1969591.1969594}.

%
% WHAT ARE THE PROBLEMS THAT WE SEE
%
By exploiting the model in practice in AOP, two main relevant issues emerged, as described below.

\paragraph{Plans without explicit goals}

% In PRS/dMARS/.. plan model, invocation conditions are meant to represent the circumstances under which the plan should be considered, usually specified in terms of events.
%

The invocation condition could be both events concerning new goals or belief changes related to percepts from the environment or data in general.
The latter case makes it possible to define reactive behaviours and data-driven / event-driven processing.
For example, the plan ``makeTea'' may be triggered by the event ``thirsty''~\cite{D'Inverno:2004:DAS:974009.974022}.

By adopting this choice in practice, a couple of important characteristics can be noted.
First, in plans triggered by invocation conditions that concern environment/data events, the state of affairs remains implicit, in the mind of the designer.
This has a drawback at runtime: intentions that are created to execute the plan are task-less/goal-less, i.e., they do not have an explicit ``state of affairs'' they are associated with.
From a design/engineering point of view, the reactive behaviour of an agent is always motivated by some task to be achieved, a ``state of affairs'' to be achieved.
In the example, the goal is to make a tea, to have a new tea. A robot that reacts to a low-battery charge event and goes back to the recharge station does so due to a maintenance task that could be described as ``keep the battery level not lower than some threshold''.
%

% This violates both the conceptual milestones, and makes it more difficult for agents to reason about what they are doing at runtime.

Second, this choice impacts strongly on modularity and reuse.
From the example, there could be multiple reasons for making a tea, not only being thirsty, and the plan -- as a recipe for making tea -- would be the same. 
However, if we put the triggering condition in the plan, that plan cannot be reused for different triggering conditions.
In this case the invocation condition appears better modelled as the motivation for adopting some specific goal (``to make a cup of tea'') and we can have different plans for it depending on the context.
But being thirsty in this case would not be part the plan itself.

\paragraph{Reactive behaviour in plan strategy}

The plan body is meant to represent a recipe for how to achieve some state of affairs.
In many relevant cases in practice, such a recipe may include the capability to asynchronously react to events from the environment.
In some cases this is about critical situations -- e.g., low battery for a robot cleaning the floor.
In some other cases, it could be strongly related to the designed strategy to fulfil the task -- e.g., reacting to messages received or monitoring some state of the ongoing work done in the environment.
In the general case, the strategy used to achieve the state of affairs may be required to flexibly mix proactive and reactive behaviour, but in the context of the same intention.

The model for plans (body) in BDI --- i.e., a sequence of actions and (sub-)goals --- does not make this straightforward.
If we need to react to some event $e$ in the context of a plan to achieve some goal $G$, then a different plan specifying $e$ as triggering condition must be used.
The effect is a poor level of \emph{encapsulation} about the plan strategy, which must be necessarily specified  in terms of a set of unrelated plans.
As in the previous case, the relation between the plans is in the mind of the designer, but is neither expressed explicitly in the source code nor is it captured by intentions at runtime.

% In AI and historically, the body of plans did not include events, but a classic process-oriented approach has been always considered.
%
% This lead to "wait for" family actions. These could be effective when from some specific point of time on we want to wait some event. 
%
% However they are not an effective way to model all the possible reactive behaviour.
%
% For instance: recharge battery case.

%In particular wait for a "synchronous" management of events. However most cases are more "asynchronous".

% Besides, they could be error-prone. 
%
% For instance, in a course of action like: \texttt{a1; a2; wait for e; a3; a4}.
%
%If the event e is happening after the action "wait for e" the semantics is clear, we know what happens: the plan is suspended until e occurs and then it goes on with a3. But what happens if e happens before the execution of "wait for e" e.g. between a1 and a2 or a2 and wait? It depends. It is a problematic case.

% So from an design/programming perspective, we see here the opportunity to improve the modularity/encapsulation of plans by allowing to specify more easily and effectively the reactive behaviour inside the body of the plan, as part of the strategy to achieve or maintain some task, which may include also further not-reactive behaviours.

\medskip

To deal with these issues, we propose an extension of the plan model adopted in BDI agents which:

\begin{itemize}

\item enforces goal/task orientation, that is:
every plan $p$ has an explicit account for the task $t$ to be either achieved or maintained.
The unique invoking condition is always a goal/task to be achieved or maintained.
This implies that every intention at runtime -- as a plan in execution -- is bound to an explicit goal.
\item extends the plan specification to include both subplans and reactive behaviour, besides a plan body having squences of actions and subgoals, so as to get full encapsulation of proactivity and reactivity in the definition of the strategy of a plan. 
\end{itemize}

\section{Proposal}
\labelsec{proposal}
%% RHB

In this section, our proposal is presented formally and independent of a particular programming language. This makes our approach more easily applicable to the various existing agent programming languages. Later in this paper we show how this has been implemented in two different agent platforms.

The presentation is formal yet language independent with the help of an abstract syntax for relevant structures such as plans, intentions, events, and so forth. The grammar below formally defines sets of such structures which can then be formally used in an algorithm of how a general BDI interpreter should be adapted to follow our approach. However, because there is no concrete syntax associated with those sets BDI structures, this makes any BDI-inspired platform that has particular representations for those structures able to use the approach in a straightforward manner. The abstract syntax is shown in Figure~\ref{fig:grammar}.

\begin{figure}[htbp]
  \footnotesize{\itshape
    \begin{center}
      \begin{tabular}{l c l l} 
   ag  & ::= &  bs gs gps \\
   bs  & ::= &  b$_{1}$ $\ldots$ b$_{n}$   & (n $\geq$ 0)\\
   gs  & ::= &  g$_{1}$ $\ldots$ g$_{n}$   & (n $\geq$ 0)\\
   ps  & ::= &  p$_{\texttt{1}}$ $\ldots$ p$_{\texttt{n}}$  & (n $\geq$ 1) \\
   p   & ::= &  g f$_{c}$ f$_{g}$ h pr$_{1}$ $\ldots$ pr$_{n}$ & (n $\geq$ 0) \\
   r   & ::= &  (+b \galt -b) f h \\
   pr  & ::= &  p \galt r\\
   t   & ::= &  g \galt +b \galt -b\\
   h   & ::= &  d$_{1}$ $\ldots$ d$_{n}$  & (n $\geq$ 0)\\
   d   & ::= &  a \galt g
\end{tabular}
    \end{center}
}
\caption{Abstract Syntax}
\label{fig:grammar}
\end{figure}

The grammar in Figure~\ref{fig:grammar} defines all structures of interest to our BDI approach, where $b$ is a metavariable standing for an individual belief an agent may have, $g$ a goal, $a$ is an action, and $f$ a logical formula.
%Events typically refer to changes in beliefs or goals the agent has adopted, but what are the exact events used and 
Note that how these and other data structures required in our presentation here are effectively expressed in a concrete programming platform is irrelevant from the abstract syntax point of view and therefore easy to adapt to whatever notion of such structures that a particular BDI programming language platform has.

Overall the grammar states that an agent is defined as a set of (initial) beliefs, a set of (initial) goals, and a set of plans. Each plan $p_i \in ps$ is a tuple that has an event, a formula often referred to as the \emph{context} or \emph{guard} for the plan (a formula checked when we are selecting a plan for an event), another formula that in~\cite{Ricci:2018:AEA:3237383.3238069} has been named the \emph{goal condition} (the goal is achieved when/if this formula becomes true given the agents beliefs). Plans $p$ and reactive rules $r$ are triggered ($t$) by a goal or changes (addition `$+$' or deletion `$-$') in beliefs, respectively. The body ($h$) of plans and rules might be empty but otherwise is sequence of deeds ($d$), a name used in~\cite{dennis08:_gwend} to refer to either goals to achieve or actions to execute. The deeds in a plan body start to be executed when an instance of that plan/rule becomes an intention. Within a plan there is a finite sequence of other such plans or reactive rules ($pr$).

We now show some algorithms that completely define our proposal in a general way for any BDI interpreter. Algorithm~\ref{alg:init} simply sets up the agent's initial state, with the help of the function for generating events in Algorithm~\ref{alg:genevents}, and all the important operations of our approach is given as an agent reasoning cycle as shown in Algorithm~\ref{alg:reascycle}. That algorithm makes reference to another algorithm used to retrieve relevant plans, shown as Algorithm~\ref{alg:getrel}, which in our approach is different than in previous agent languages, because the plans give a context in which to look for relevant plans and reactive rules.

Note that the non-terminals of the grammar define syntactic categories, for example $bs$ is a set of beliefs, and when needed we will index its elements by natural numbers. For example, given an agent  $ag = \langle bs, gs, ps \rangle$, we may refer to $b_i \in bs$ as the agent's $i$-th initial belief. In the algorithms introduced below, if a variable name coincides with a syntactical category of the grammar in Figure~\ref{fig:grammar}, that clearly defines its type. Otherwise, we use the syntactic categories to express the type of a variable (or indeed to check whether the content of a variable) using the notation $u : c$ to say that the content of variable $u$ conforms to the syntactic category $c$. So, for example, if we say $B : bs$ we mean that variable $B$ contains a set of beliefs.

Besides the data structures defined in the grammar, we need structures for events and intentions. An intention $i$ is a stack of plans, each with its own stack of deeds (those that are internal to plan or coming from activated reactive rules), denoted as $[p_1[d_{1_{1}},\ldots,d_{1_{n1}}],\ldots,p_n[d_{n_{1}},\ldots,d_{n_{nk}}]]$ where $p_1$ is the plan at the top of that intention stack and $d_{1_{1}}$ is the deed at the top of $p_1$'s stack of deeds (and therefore the next deed to be executed). Events typically refer to changes in beliefs or goals the agent has adopted, so an event $e$ is a tuple $\langle t, i \rangle$ where $t$ is a trigger (as defined in the grammar above) and $i$ an intention, possibly~$[]$, the empty intention. As in some agent languages, when an event is associated with an empty intention we call it \emph{external}; external events arise from perception of the environment, agent communication (for example, goal delegation) or initial goals, whereas internal events are associated with an intention (for example, a plan being executed requires the achievement of a subgoal).

The main agent algorithm is shown in Algorithm~\ref{alg:init}, which simply initialises variable $B$ for the agent beliefs, $P$ for the agent plans, and the set of events $E$ initially contains one event for each initial goal in the program (if any). The set of intentions $I$ is initially empty. In this presentation an agent enters an infinite loop within which it: (i) perceives the environment with a function $\Call{current\_percepts}{}$ which is assumed as given (i.e., a concrete agent architecture will have the ability to do so); (ii) generate events for all \emph{changes} in beliefs caused by the received percepts (see Algorithm~\ref{alg:genevents}); (iii) uses the $\Call{reasoning\_cycle}{}$ function shown in Algorithm~\ref{alg:reascycle} to decide on the next action to take; an (iv) executes that action (again it is assumed that the actual agent architecture provides the means for executing an action, which expressed in the algorithm by function $\Call{execute}{}$).

\begin{algorithm}
  \caption{Agent Initialisation
    \label{alg:init}}
  \begin{algorithmic}[1]
    \Require{an initial agent program $ag = \langle bs, gs, ps \rangle$}
    \State $B \gets bs$
    \State $E \gets \{\}$
    \ForAll{$g \in gs$}
        \State $E \gets E \cup \{ \langle g, [] \rangle \}$
    \EndFor
    \State $P \gets ps$
    \State $I \gets \{\}$
    \While{$true$}
        \State $S \gets \Call{current\_percepts}{}$
        \State \Call{generate\_events}{S}
        \State $action \gets \Call{reasoning\_cycle}{~}$
        \State \Call{execute}{action}
    \EndWhile
  \end{algorithmic}
\end{algorithm}

Algorithm~\ref{alg:genevents} simply checks whether there are received percepts (i.e, symbolic information saying what is currently perceived as true in the environment) which are not currently in the belief base and generate external events for the addition of those beliefs (line~\ref{line:addbel}). It then checks if there are beliefs which are no longer perceived as true in the environment and generates belief deletions events for those (line~\ref{line:delbel}). The belief base itself is updating accordingly.

\begin{algorithm}
    \caption{Event Generation from Percepts
    \label{alg:genevents}}
    \begin{algorithmic}[1]
        \Require{external variables $B, E$}
        \Function{generate\_events}{$S$}
%            \Statex
            \State \Comment{$S$ is a set of percepts from the agent's sensors\hspace*{1.4cm}~}
            \ForAll{$s \in S$}
                \If{$s \not\in B$}
                    \State $B \gets B \cup \{s\}$
                    \State $E \gets E \cup \{ \langle +s, [] \rangle \}$\label{line:addbel}
                \EndIf
            \EndFor
            \ForAll{$b \in B$}
                \If{$b \not\in S$}
                    \State $B \gets B \setminus \{b\}$
                    \State $E \gets E \cup \{ \langle -b, [] \rangle \}$\label{line:delbel}
                \EndIf
            \EndFor
        \EndFunction
    \end{algorithmic}
\end{algorithm}

%%RHB: Here, explain get_applicable, select_*, etc...
The main algorithm is in fact Algorithm~\ref{alg:reascycle}, which shows the reasoning cycle for a BDI agent following our approach. Recall that in our approach we have a ``goal condition'' which, when true, implies that a corresponding intended means can be dropped. This leads to some computational burden but it is an important feature as it ensures the agent will not take unnecessary action. This is done in lines \ref{line:drop1}--\ref{line:drop2}, starting from the bottom of the intention because dropping a goal near the bottom, if possible, will eliminate all the stack of plans on top of it within the intention.

We then handle one event from the set of events $E$. Note that, as in AgentSpeak, we assume there are user-defined functions to select an event, an option or intended means (one among possibly various applicable plans), and an intention from the set of intentions to execute next, respectively called in the algorithm $\Call{select\_ev}{}, \Call{select\_opt}{},$ and $\Call{select\_int}{}$. Function~\Call{get\_applicable}{} is not given an algorithm because it simply checks if the context part of the plans or rules are true (depending on the particular language, this might include checking for logical consequence from the belief base, for example).  After one event is selected, there are different parts of the algorithm depending on what type of event was selected.

For a belief-change event, this is handled in lines \ref{line:rr1}--\ref{line:rr2}. Differently from other BDI platforms, our approach will replicate the reactive event to each existing intention for which it is relevant. So for each intention in the set of intention, we search for relevant plans in the entire intention, which in our notation here includes syntactical copies of the plan being executed, including therefore the plans encapsulated in them. If there are relevant and applicable plans for that particular intention, one of those is selected and its body executed. Again, we emphasise this process is repeated for each single intention.

External goal events, i.e., initial goals or goals delegated by other agents, are handled in lines~\ref{line:eg1}--\ref{line:eg2}. For those, after selecting one applicable relevant plan, we simply create a new intention for it if the goal condition is not yet believed true; the intention is a stack with a single element which has a copy of the plan $p$ and a list of deeds to be executed (the body of that plan).

Internal goal events, i.e., subgoals that appear in currently executing plans, are handled in lines~\ref{line:ig1}--\ref{line:ig2}. This is very similar to the case above except that new the intended means is pushed on top of the existing intention.

The final part of the algorithm selects one intention to be further executed in that reasoning cycle. If the first deed not yet executed in the body of the topmost plan in the intention is a goal, the appropriate event is generated and the reasoning cycle function called recursively until a next action to execute is determined. If the deed is an action, the intention is updated to reflect that the action is going to be executed (i.e., it is removed of the list of deeds to be executed) and the action is simply returned.

\begin{algorithm}
    \caption{Reasoning Cycle
    \label{alg:reascycle}}
    \begin{algorithmic}[1]
        \Require{external variables $B, E, P, I$}
        \Function{reasoning\_cycle}{}
            \State \Comment{Drop all achieved goals\hspace*{4.3cm}~}
            \ForAll{$i \in I$}\label{line:drop1}
                \For{$j = \Call{length}{i}$ down to $1$} \Comment{From bottom to top}
                    \State $i'' \gets [p_j[h_j]] \in i$
                    \State let $p_j = \langle t,f_c,f_g,h,pr_1,\ldots,pr_n\rangle$
                    \If{$B \models f_g$}
                        \State $I \gets I \setminus \{i\} \cup \{ \Call{remove}{i'',i} \}$
                    \EndIf
                \EndFor
            \EndFor\label{line:drop2}
            \State \Comment{Handle an event\hspace*{5.2cm}~}
            \State $se \gets \Call{select\_ev}{E}$\
            \State let $se = \langle t, i \rangle$
            \If{$i=\{\} \wedge t : (+b \galt -b)$} \Comment{external belief event}
                \ForAll{$i \in I$} \Comment{may trigger a r.r. in each intention}\label{line:rr1}
                    \State $rps = \Call{get\_relevant}{t,i,\{\}}$ \Comment{search entire $i$}
                    \State $aps = \Call{get\_applicable}{rps}$
                    \If{$aps \neq \{\}$}
                        \State $r = \Call{select\_opt}{aps}$
                        \State let $r = \langle t f h\rangle$
                        \State let $i = [p_1[h_1],i']$ 
                        \State $i'' \gets [p_1[h,h_1],i']$
                        \State $I \gets I \setminus \{i\} \cup \{i''\}$
                    \EndIf
                \EndFor\label{line:rr2}
            \ElsIf{$i=\{\} \wedge t : g$} \Comment{external goal event}
                \State $rps \gets \Call{get\_relevant}{t,[],\{\}}$ \Comment{search only in P} \label{line:eg1}
                \State $aps \gets \Call{get\_applicable}{rps}$
                \If{$aps \neq \{\}$}
                    \State $p \gets \Call{select\_opt}{aps}$
                    \State let $p = \langle t,f_c,f_g,h,pr_1,\ldots,pr_n\rangle$
                    \If{$B \not\models f_g$}
                        \State $I \gets I \cup \{[p[h]]\}$
                    \EndIf
                \EndIf\label{line:eg2}
            \ElsIf{$i\neq\{\} \wedge t : g$} \Comment{internal (goal) event}
                \State $rps \gets \Call{get\_relevant}{t,i,\{\}}$ \Comment{search in $i$} \label{line:ig1}
                \State \Comment{and top level plans in $P$}
                \State $aps \gets \Call{get\_applicable}{rps}$
                \If{$aps \neq \{\}$}
                    \State $p \gets \Call{select\_opt}{aps}$
                    \State let $p = \langle t,f_c,f_g,h,pr_1,\ldots,pr_n\rangle$
                    \If{$B \not\models f_g$}
                        \State $I \gets I \setminus \{i\} \cup \{\Call{push}{p[h],i}\}$
                    \EndIf
                \EndIf\label{line:ig2}
            \EndIf
            \State \Comment{Execute a step of an intention\hspace*{3.4cm}~}
            \State $i \gets \Call{select\_int}{I}$
            \State let $i = [p_1[d_1,h_1],i']$
%            \State $d \gets \Call{get\_first}{}(\Call{get\_top}{i})$
            \If{$d_1 : g$}
                \State $E \gets E \cup \{\langle d_1,i\rangle\}$
                \State $action \gets \Call{reasoning\_cycle}{~}$
                \State \Return{$action$}
            \ElsIf{$d_1 : a$}
                \State $i'' \gets [p_1[h_1],i']$
                \State $I \gets I \setminus \{i\} \cup \{i''\}$
                \State \Return{$d_1$}
            \EndIf
            \State \Comment{See note in the text about empty stacks\hspace*{2.3cm}~}
        \EndFunction
    \end{algorithmic}
\end{algorithm}

Note that in Algorithm~\ref{alg:reascycle} we do not give the details of what happens when a stack of deeds associated with a plan in the intention stack becomes empty. In practice that leads to the removal of that plan from the intention stack and possibly the subgoal in the list of deeds of the plan below it in the intention stack (or the entire intention if the list of deeds at the bottom is then empty).

\begin{algorithm}
    \caption{Retrieving Relevant Plans
    \label{alg:getrel}}
    \begin{algorithmic}[1]
        \Require{external variable $P$}
        \Function{get\_relevant}{t,i,rps}
%            \State $rp \gets \{\}$
%            \For{$j = 1$ to $\Call{length}{i}$} \Comment{From top to bottom}
            \If{$i \neq \{\}$}
                \State let $p[h] = \Call{head}{i}$
                \ForAll{$pr \in p$}
                    \If{\Call{relevant}{pr,t}}
                        \State $rps \gets rps \cup \{pr\}$
                    \EndIf
                \EndFor
                \Call{get\_relevant}{}$(t,\Call{tail}{i},rps)$
%                \If{$rp \neq \{\}$}
%                    \State \Return{$rp$}
%                \EndIf
            \Else
                \ForAll{$p \in P$} %\Comment{none in the intention scope}
            \Comment{check for relevant top-level plans}
                    \If{\Call{relevant}{p,t}}
                        \State $rps \gets rps \cup \{p\}$
                    \EndIf
                \EndFor
            \EndIf
            \State \Return{$rps$}
        \EndFunction
    \end{algorithmic}
\end{algorithm}

Finally, Algorithm~\ref{alg:getrel} shows a recursive function which receives as parameter a trigger $t$ to be matched with the triggers of plans anywhere in the intention $i$ and also in top-level goals of the plan library (note that in our approach, reactive rules do not appear at the top-level of a plan library, only associated with the goals that the agent may need to achieve). The final parameter is a set of plans already determined to be relevant for $t$, normally empty when the function is initially called. Function \Call{relevant}{} simply tries to match the triggering-event of plan $p$ with $t$ (the implementation of this function is of course language dependent; it might require for example unification if the language is logic based).

\section{Implementation and Evaluation}
\labelsec{implementation}

To assess the computational viability and characteristics of the proposed language, we extended the interpreters of both ASTRA~\cite{DBLP:conf/prima/CollierRL15} and Jason~\cite{bordini:07}, called ASTRA(ER) and Jason(ER) respectively, with the new features discussed in this paper. The implementation has helped us not only confirm that the ideas are feasible, but provided refinements for the model, as well as being a tool to measure how the approach scales and compares against other languages.\footnote{ASTRA(ER) is available at \url{xxxxxxxx} 
% https://tinyurl.com/rsrm2eu
and Jason(ER) is available at \url{xxxxxxxx}.}
% https://tinyurl.com/vru3mw5

\subsection{Jason(ER)}

An example of a concrete program in Jason(ER) is shown in Listing~\ref{alg:jasoner}. It implements plans for an initiator agent in the context of the Contract Net Protocols (CNP)~\cite{fipa29:00}. The main part of the program is a plan to achieve the goal of running a CNP for some given task (lines 4--14). This plan encapsulates a body (line 5) and three sub-plans (lines 6--13). The body has three sub-goals: announce the call for proposals (CFP), wait for bids, and contract the winner. We highlight here the sub-plan for the sub-goal \texttt{bids}. It is considered achieved when either ($i$) all participants have sent an answer (a proposal or a refusal) or ($ii$) a deadline has passed (four seconds in this case). The rule on line~1 is used to evaluate condition ($i$) and the \texttt{.wait} on line~7 is used for condition ($ii$).

The challenge is on the interplay of those two conditions. The solution here uses two features proposed in this paper: goal conditions and sub-plans and reactive rules. The goal condition is placed after \texttt{<:} on line~6 and is \texttt{false}. Since this condition is never satisfied in the mental state of the agent, the internal action \texttt{.done} is used to finish the goal. The reactive rules on lines~9 and~10 are used to react to the answers. Thus, while running line~7 of the plan body, for every received answer, one of those two reactive rules is selected and, if enough answers have being received, it finishes the goal \texttt{bids}. If none of these two rules are executed, the plan body executes \texttt{.done} after four seconds, continuing with the sequence on line~7. 

Considering this program, we emphasise that:
\begin{itemize}
    \item The plan to achieve goal \texttt{bids} \emph{encapsulates} both proactive and reactive behaviour. The former as a sequence of actions (the body) and the latter as a set of (encapsulated) reactive rules.
    \item The reaction to answers is defined in the context of the goals \texttt{bids} and \texttt{cnp}. An agent knows therefore why (for which goal) it is executing those reactive rules. 
\end{itemize}

\begin{listing}
  \caption{Jason(ER) implementation of CNP initiator} \label{alg:jasoner}
%\begin{minted}[linenos]{prolog}
\small{
\begin{verbatim}
all_ans(I) :- ... // true if all participants have answered

// plan to achieve goal cnp, I identifies the CNP
+!cnp(I,Task) {
    <- announce_cfp(I,Task); !bids(I); !contract(I). 
    +!bids(I) <: false {
        <- .wait(4000); .done.
        // reaction to the event of new proposal / refusal
        +propose(I,_) : all_ans(I) <- .done.
        +refuse(I)    : all_ans(I) <- .done.
    }
    +!announce_cfp(I,Task) <- ...
    +!contract(I) <- ...
}
\end{verbatim}}
%\end{minted}
\end{listing}

\begin{figure}
    \centering
    \includegraphics[width=\linewidth]{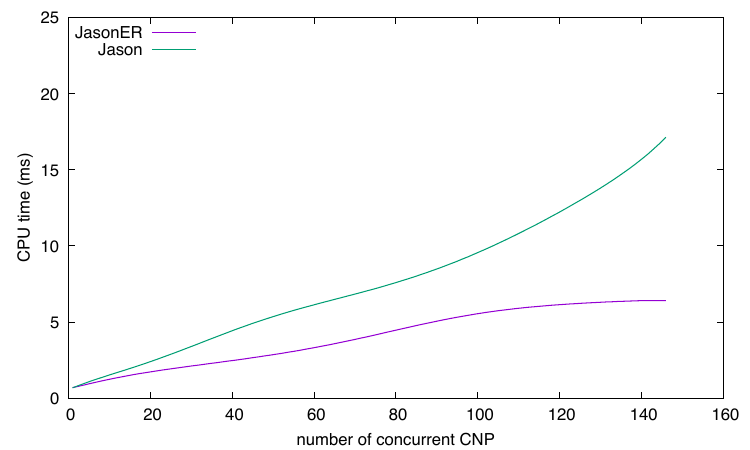}
    \caption{Jason(ER) performance evaluation.}
    \label{fig:jasonertime}
\end{figure}

The implementation of Jason(ER) allowed us to evaluated how it scales considering a MAS that concurrently runs $n$ CNPs. It is expected that the time required to finish $n$ CNPs increases linearly on $n$. The MAS has one agent playing initiator and eleven playing participant. Only the initiator uses the new features of Jason(ER) as shown in Listing~\ref{alg:jasoner}. The result of the experiment, shown in Figure~\ref{fig:jasonertime}, confirms that Jason(ER) scales linearly on the number of CNPs.%\footnote{The code and data required to repeat the experiment, as well as more details, are available at \url{xxxxxxxx}.}
% https://tinyurl.com/qwamg93

In a second experiment, we intend to evaluate the overhead of the new features. We thus run the same MAS replacing the initiator agent by one implemented as usual in Jason (its program is shown in Listing~\ref{alg:jason}). We noted that Jason(ER) is indeed faster than Jason for this application (cf.\ Figure~\ref{fig:jasonertime}). The reasons for this old Jason agent being slower are the following:
\begin{enumerate}
\item 
While sending the CFP and before finishing this task, the agent starts receiving proposals and refusals for the first sent messages. This agent has relevant plans for these events (lines 5--6) even if (for sure) these plans are not applicable yet --- they can be applicable only after the last CFP was sent. The agent is thus wasting time considering the plans on lines~5 and~6 before  finishing the CFP announcements.
\item
The contract goal (line~9) needs to be singleton for a CNP (we cannot contract twice for the same CNP), however plans on lines~5 and~6 may trigger this goal more than once. Thus the context of this plan tests if no other contract intention is running for the same CNP; the internal action \texttt{.intend} performs this test. This internal action complexity is $O(m)$ (where $m$ is the number of active intentions). Jason(ER) avoids these two issues: plans for proposals and refusals are relevant only when the goal \texttt{bids} is being pursued (neither before, nor after); the goal \texttt{contract} is pursued after the goal \texttt{bids} and not as a consequence of a reaction.
\end{enumerate}

Despite the improvements in the performance (that could be circumstantial for this very experiment/application), the main advantage of Jason(ER) is how long-term goals (line~5 in Listing~\ref{alg:jasoner}) are explicitly related to relevant reactions (lines 9--10). The decomposition into sub-goals is also more explicit, as we can see when comparing line~5 in Listing~\ref{alg:jasoner} against line~1 in Listing~\ref{alg:jason}. While in the former the interpreter can foresee \texttt{contract} as a potential future goal, in the latter, it cannot.

\begin{listing}
  \caption{Jason implementation of CNP initiator} \label{alg:jason}
%\begin{minted}[linenos]{prolog}
\small{
\begin{verbatim}
+!cnp(I,Task) <- announce_cfp(I,Task); !bids(I).

+!bids(I) <- .at("now +4 seconds", { +!contract(I) }).

+propose(I,_) : all_ans(I) <- !contract(I).
+refuse(I)    : all_ans(I) <- !contract(I).

+!announce_cfp(I,Task) <- ...
+!contract(I) : not .intend(contract(I)) <- ...
\end{verbatim}}
%\end{minted}
\end{listing}

\subsection{ASTRA(ER)}

\begin{figure}
    \centering
    \includegraphics[width=\linewidth]{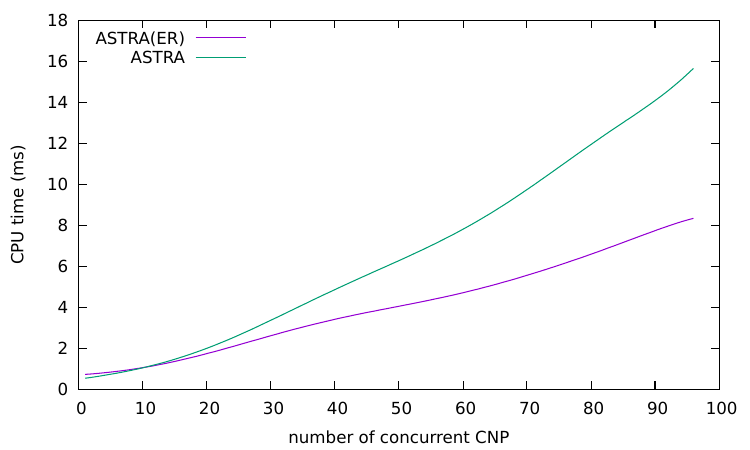}
    \caption{ASTRA(ER) performance evaluation.}
    \label{fig:asertime}
\end{figure}

A similar experiment was run comparing ASTRA and ASTRA(ER) using the CNP benchmark. The results of this experiment are presented in figure \ref{fig:asertime}. They demonstrate that ASTRA(ER) also outperforms ASTRA. This is very encouraging given that, unlike the Jason code, the two ASTRA versions are virtually identical. The main difference is the use of encapsulation of plans.

In addition to studying the performance of our new language, it is also interesting to consider the effect that encapsulated plans have on the design of larger programs.
%
% we can remove this if necessary...
We have seen that protocols, such as CNP, benefit from the use of encapsulated plans. But, are there other areas that can benefit? One such area we have identified is queue/list processing. There are many scenarios in which an agent needs to process the items contained within a list --- for example, a list could contain an agreed list of tasks to be executed by the agent. The code for such a scenario is presented in Listing~\ref{alg:processor}. 
This listing contains two agent programs --- an encapsulated plan for processing queues (\verb|QueueProcessor|) and a second program \verb|Main| which is a somewhat contrived example that uses this encapsulated plan to execute a set of purchases and payments.

\begin{listing}
  \caption{ASTRA implementation of a Queue Processor} \label{alg:processor}
%\begin{minted}[linenos]{prolog}
\small{
\begin{verbatim}
agent QueueProcessor {
  module System system;
  goal +!processQueue(list queue, int delay) {
    body { !processQueueItems(queue); }
    rule +!processQueueItems([funct I | list T]) {
      !processItem(I);
      if (list_count(T) > 0) {
        system.sleep(delay);
        !processQueueItems(T);
      }
    }
  }
  rule +!processQueueItem(funct I) {system.fail();}
}
agent Main extends QueueProcessor {
  initial !performTasks([buy("apples"), pay(2)]);
  goal +!performTasks(list events) {
    body {
      !processQueue(events, 3000);
    }
    rule +!processItem(buy(string I)) {
      console.println("buying: "+ I);
    }
    rule +!processItem(pay(int amt)) {
      console.println("paying: " + amt);
    }
  }
}
\end{verbatim}}
%\end{minted}
\end{listing}

What is interesting to note in this example is the use (and otherwise) of encapsulated plans. The \verb|!processQueue| plan includes a plan to handle the recursive processing of the queue (including a specified delay between processing steps). However, the plan to handle the processing of an individual item is outside the goal plan. This has been done intentionally --- if the plan for processing items was encapsulated within the goal rule, then it would always be given precedence over any other rules as our algorithm checks plans in the intention before it checks plans outside of the intention. This has the effect of making the behaviour immutable, in that it can never be overridden by other code.
By declaring the plan outside of the goal plan, the plan becomes mutable --- we can override the plan in any agent program that extends the \verb|QueueProcessor| program.

This is precisely what the second agent program in listing~\ref{alg:processor} does. Here, we see that there is a single goal plan that declares a subgoal that can be handled by the \verb|!processQueue| plan. However, in this case, the \verb|!performTasks| goal plan also includes encapsulated rules for handling the specific types of task that are given in the input queue. The point here is that we are able to define a mutable templated behaviour (the \verb|QueueProcessor|) which can be included in different programs where the expected behaviour is defined immutably (so how the \verb|Main| program responds to the \verb|buy(...)| and \verb|pay(...)| tasks can never be changed). However, the program can be further extended to support additional tasks as is required.

This type of support speaks directly to long cited needs for libraries of reusable agent behaviours. While the implementation of templated behaviour was possible in AgentSpeak(L), they were reliant on hidden state and could easily be modified to work in unintended ways.

In order to better illustrate the potential of the approach, we refer the reader to a larger example that is available online. This example implements a one-shot first price auction infrastructure using a combination of the FIPA request, subscribe, and CNP protocols together with the \verb|QueueProcessor| example shown above which is used to simulate expected interactions with the system.

\section{Related Work}
\labelsec{related}

Several implementations and extensions of the BDI model have been proposed in literature, since the PRS~\cite{Ingrand:1992:ARR:629535.629890},  dMARS~\cite{D'Inverno:2004:DAS:974009.974022}  and AgentSpeak(L)~\cite{Rao96}.
Several agent programming languages and development platforms in the BDI tradition are available, besides Jason~\cite{jason06,bordini:07} and ASTRA~\cite{DBLP:conf/prima/CollierRL15} used in this paper.
Main examples are JADEX~\cite{DBLP:books/sp/map2005/PokahrBL05}, 3APL~\cite{Hindriks:1999:AP:608601.608641}  and 2APL~\cite{Dastani2008}, GOAL~\cite{Hindriks:09a}, Jack~\cite{DBLP:books/sp/map2005/Winikoff05}, SRI’s SPARK~\cite{Morley:2004:SAF:1018410.1018821}, and JAM~\cite{Huber:1999:JBM:301136.301202}.
Among the most recent ones, we mention CANplan~\cite{Sardina:2006:HPB:1160633.1160813}, an agent-oriented programming language that enhances usual BDI programming style with three distinguished features, namely declarative goals, look-ahead planning, and failure handling.
All existing implementations preserve the original elements of the BDI plan model based on plans where the triggering conditions can be events about the environment and the plan body is a course of action including sub-goals and primitive actions.
\todo{GO DEEPER - ADDING SOME DETAILED COMPARTISON}

%
% ASER
%
As far as authors' knowledge, AgentSpeak(ER)~\cite{Ricci:2018:AEA:3237383.3238069} is the proposal extending the plan model in the direction discussed in this work.
This paper generalises and extends the 
%seminal 
model proposed in~\cite{Ricci:2018:AEA:3237383.3238069}, providing both a formalisation and a concrete implementation and evaluation based on two different platforms, namely Jason and ASTRA.

%
% AOP in GENERAL
%
Devising effective programming models for integrating proactive and reactive behaviour is a general aim of Agent-Oriented Programming~\cite{DBLP:journals/ai/Shoham93} and different approaches can be found in literature  proposed for agent programming languages,  not strictly based on BDI.
%

%%RHB: As I don't know what to write here, I'm commenting it out in case we do a last minute submission
\begin{comment}
\begin{itemize}
\item Teleo-reactive~\cite{Nilsson:1994:TPA:1618595.1618602} and TeleoR~\cite{7139900} 
\item logic-based: GOLOG family (GOLOG~\cite{LEVESQUE199759}, ConGOLOG~\cite{DEGIACOMO2000109}, IndiGolog~\cite{DeGiacomo2009})
\end{itemize}
%
\todo{COMMENTING}
\end{comment}

% Relationship with /impact on the IPP~\cite{Logan:2017:PIP:3091125.3091234}.

%
% MODULARITY
% 
Finally, encapsulation is strongly related to modularity, typically impacting on how (agent) programs are organised.
So this work is related to contributions in the literature focusing on improving modularity in BDI agent programming~\cite{Madden2010,Busetta2000,5285116,Novak:2006:MBA:1160633.1160814,vanRiemsdijk:2006:GMA:1160633.1160864,Ortiz-Hernandez2016,Hindriks2008,Nunes2014,10.1007/11678823_9}.
In that literature, modules are typically used as a mechanism to structure agent programs in separate parts (modules), each encapsulating cognitive components such as beliefs, goals, and plans that together model a specific functionality and can be
used to handle specific situations or tasks~\cite{5285116}.
%
% In programming languages, modularity is strongly related to encapsulation. 
%
% Strengthening encapsulation typically leads to refining modularity, in particular devising more coarse-grained modules.
%
%
%
From a software engineering point of view, modules allow a programmer to focus on those skills that are required to handle a situation~\cite{Hindriks2008}. 
In that perspective, the approach proposed in this paper improves modularity by devising coarse-grained plans encapsulating goal-oriented \emph{and} reactive behaviour. 
This approach can be integrated with existing more comprehensive proposal about modularity as \cite{Madden2010}, where a module is  meant to be a composable subset of the functionality of an agent, represented by a functional unit encapsulating goals, beliefs, and plans.
\todo{TO BE EXTENDED}

\section{Conclusions}
\labelsec{conclusions}

In this paper, we have put forward a new model for plans in BDI agent programming that provides encapsulation of both proactive and reactive behaviour. We have given algorithms for how that model is to be used in a BDI agent reasoning cycle in a way that is abstract enough to be useful for implementation in existing and future BDI platforms. That is, the algorithms are based on abstract syntax so that the approach is language independent. Furthermore, we have implemented these algorithms in two different platforms, namely Jason and ASTRA. Finally, we have reported on experimental results showing that at least in some applications our approach leads to significant efficiency compared to the traditional approach. More importantly, our examples show the elegance of our model when instantiated in particular languages and in fact some of the advantages for programming practice.

Future work includes implementing the model in other languages and further evaluating its performance. More importantly, we aim to develop more complex systems with the extended languages in order to fully assess the practical impact of our approach on agent development.

\newpage

\bibliographystyle{ACM-Reference-Format}
\bibliography{references}

@manual{fipa29:00,
	Address = {Geneva, Switzerland},
	Author = {Foundation for Intelligent Physical Agents},
	Institution = {Foundation for Intelligent Physical Agents},
	Title = {{FIPA} Contract Net Interaction Protocol},
	Url = {http://www.fipa.org},
	Year = {2000}}

@book{DBLP:books/sp/map05,
  editor    = {Bordini, Rafael H. and
               Mehdi Dastani and
               J{\"{u}}rgen Dix and
               Amal El Fallah{-}Seghrouchni},
  title     = {Multi-Agent Programming: Languages, Platforms and Applications},
  series    = {Multiagent Systems, Artificial Societies, and Simulated Organizations},
  volume    = {15},
  publisher = {Springer},
  year      = {2005},
  isbn      = {0-387-24568-5},
  bibsource = {dblp computer science bibliography, https://dblp.org}
}

@book{DBLP:books/sp/FDDB2009,
  editor    = {Bordini, Rafael H. and
               Mehdi Dastani and
               J{\"{u}}rgen Dix and
	       Amal El Fallah{-}Seghrouchni},
  title     = {Multi-Agent Programming, Languages, Tools and Applications},
  publisher = {Springer},
  year      = {2009},
  url       = {https://link.springer.com/book/10.1007/978-0-387-89299-3},
  isbn      = {978-0-387-89298-6},
  bibsource = {dblp computer science bibliography, https://dblp.org}
}

@InProceedings{dennis08:_gwend,
  author = 	 {Louise A. Dennis and Berndt Farwer},
  title = 	 {Gwendolen: A BDI Language for Verifiable Agents},
  booktitle = 	 {Logic and the Simulation of Interaction and Reasoning},
  year =	 2008,
  editor =	 {Benedikt L{\"o}we},
  address =	 {Aberdeen},
  publisher =	 {AISB},
  note =	 {AISB'08 Workshop}
}

@inproceedings{Ricci:2018:AEA:3237383.3238069,
 author = {Ricci, Alessandro and Bordini, Rafael H. and H\"{u}bner, Jomi F. and Collier, Rem},
 title = {AgentSpeak(ER): An Extension of AgentSpeak(L) Improving Encapsulation and Reasoning About Goals},
 booktitle = {Proceedings of the 17th International Conference on Autonomous Agents and MultiAgent Systems},
 series = {AAMAS '18},
 year = {2018},
 location = {Stockholm, Sweden},
 pages = {2054--2056},
 numpages = {3},
 url = {http://dl.acm.org/citation.cfm?id=3237383.3238069},
 acmid = {3238069},
 publisher = {International Foundation for Autonomous Agents and Multiagent Systems},
 address = {Richland, SC},
}

@INPROCEEDINGS{7139900,
author={K. L. {Clark} and P. J. {Robinson}},
booktitle={2015 IEEE International Conference on Robotics and Automation (ICRA)},
title={Robotic agent programming in TeleoR},
year={2015},
volume={},
number={},
pages={5040-5047},
doi={10.1109/ICRA.2015.7139900},
ISSN={},
month={May},}

@inproceedings{Logan:2017:PIP:3091125.3091234,
 author = {Logan, Brian and Thangarajah, John and Yorke-Smith, Neil},
 title = {Progressing Intention Progression: A Call for a Goal-Plan Tree Contest},
 booktitle = {Proceedings of the 16th Conference on Autonomous Agents and MultiAgent Systems},
 series = {AAMAS '17},
 year = {2017},
 location = {S\&\#227;o Paulo, Brazil},
 pages = {768--772},
 numpages = {5},
 url = {http://dl.acm.org/citation.cfm?id=3091125.3091234},
 acmid = {3091234},
 publisher = {International Foundation for Autonomous Agents and Multiagent Systems},
 address = {Richland, SC},
}

@article{Nilsson:1994:TPA:1618595.1618602,
 author = {Nilsson, Nils J.},
 title = {Teleo-reactive Programs for Agent Control},
 journal = {J. Artif. Int. Res.},
 issue_date = {August 1993},
 volume = {1},
 number = {1},
 month = jan,
 year = {1994},
 issn = {1076-9757},
 pages = {139--158},
 numpages = {20},
 url = {http://dl.acm.org/citation.cfm?id=1618595.1618602},
 acmid = {1618602},
 publisher = {AI Access Foundation},
 address = {USA},
}

@article{DEGIACOMO2000109,
title = "ConGolog, a concurrent programming language based on the situation calculus",
journal = "Artificial Intelligence",
volume = "121",
number = "1",
pages = "109 - 169",
year = "2000",
issn = "0004-3702",
doi = "https://doi.org/10.1016/S0004-3702(00)00031-X",
url = "http://www.sciencedirect.com/science/article/pii/S000437020000031X",
author = "Giuseppe De Giacomo and Yves LespÃ©rance and Hector J. Levesque"
}

@Inbook{DeGiacomo2009,
author="De Giacomo, Giuseppe
and Lesp{\'e}rance, Yves
and Levesque, Hector J.
and Sardina, Sebastian",
editor="El Fallah Seghrouchni, Amal
and Dix, J{\"u}rgen
and Dastani, Mehdi
and Bordini, Rafael H.",
title="IndiGolog: A High-Level Programming Language for Embedded Reasoning Agents",
bookTitle="Multi-Agent Programming: Languages, Tools and Applications",
year="2009",
publisher="Springer US",
address="Boston, MA",
pages="31--72",
isbn="978-0-387-89299-3",
doi="10.1007/978-0-387-89299-3_2",
url="https://doi.org/10.1007/978-0-387-89299-3_2"
}

@article{LEVESQUE199759,
title = "GOLOG: A logic programming language for dynamic domains",
journal = "The Journal of Logic Programming",
volume = "31",
number = "1",
pages = "59 - 83",
year = "1997",
note = "Reasoning about Action and Change",
issn = "0743-1066",
doi = "https://doi.org/10.1016/S0743-1066(96)00121-5",
url = "http://www.sciencedirect.com/science/article/pii/S0743106696001215",
author = "Hector J. Levesque and Raymond Reiter and Yves LespÃ©rance and Fangzhen Lin and Richard B. Scherl"
}

@article{DBLP:journals/ai/Shoham93,
  author    = {Yoav Shoham},
  title     = {Agent-Oriented Programming},
  journal   = {Artif. Intell.},
  volume    = {60},
  number    = {1},
  pages     = {51--92},
  year      = {1993},
  url       = {https://doi.org/10.1016/0004-3702(93)90034-9},
  doi       = {10.1016/0004-3702(93)90034-9},
  bibsource = {dblp computer science bibliography, https://dblp.org}
}

@article{Ingrand:1992:ARR:629535.629890,
 author = {Ingrand, Francois F. and Georgeff, Michael P. and Rao, Anand S.},
 title = {An Architecture for Real-Time Reasoning and System Control},
 journal = {IEEE Expert: Intelligent Systems and Their Applications},
 issue_date = {December 1992},
 volume = {7},
 number = {6},
 month = dec,
 year = {1992},
 issn = {0885-9000},
 pages = {34--44},
 numpages = {11},
 url = {http://dx.doi.org/10.1109/64.180407},
 doi = {10.1109/64.180407},
 acmid = {629890},
 publisher = {IEEE Educational Activities Department},
 address = {Piscataway, NJ, USA},
}

@article{Sardina:2011:BAP:1969591.1969594,
 author = {Sardina, Sebastian and Padgham, Lin},
 title = {A BDI Agent Programming Language with Failure Handling, Declarative Goals, and Planning},
 journal = {Autonomous Agents and Multi-Agent Systems},
 issue_date = {July      2011},
 volume = {23},
 number = {1},
 month = jul,
 year = {2011},
 issn = {1387-2532},
 pages = {18--70},
 numpages = {53},
 url = {http://dx.doi.org/10.1007/s10458-010-9130-9},
 doi = {10.1007/s10458-010-9130-9},
 acmid = {1969594},
 publisher = {Kluwer Academic Publishers},
 address = {Hingham, MA, USA},
}

@inproceedings{Sardina:2006:HPB:1160633.1160813,
 author = {Sardina, Sebastian and de Silva, Lavindra and Padgham, Lin},
 title = {Hierarchical Planning in BDI Agent Programming Languages: A Formal Approach},
 booktitle = {Proceedings of the Fifth International Joint Conference on Autonomous Agents and Multiagent Systems},
 series = {AAMAS '06},
 year = {2006},
 isbn = {1-59593-303-4},
 location = {Hakodate, Japan},
 pages = {1001--1008},
 numpages = {8},
 url = {http://doi.acm.org/10.1145/1160633.1160813},
 doi = {10.1145/1160633.1160813},
 acmid = {1160813},
 publisher = {ACM},
 address = {New York, NY, USA},
}

@inproceedings{Huber:1999:JBM:301136.301202,
 author = {Huber, Marcus J.},
 title = {JAM: A BDI-theoretic Mobile Agent Architecture},
 booktitle = {Proceedings of the Third Annual Conference on Autonomous Agents},
 series = {AGENTS '99},
 year = {1999},
 isbn = {1-58113-066-X},
 location = {Seattle, Washington, USA},
 pages = {236--243},
 numpages = {8},
 url = {http://doi.acm.org/10.1145/301136.301202},
 doi = {10.1145/301136.301202},
 acmid = {301202},
 publisher = {ACM},
 address = {New York, NY, USA},
}

@inproceedings{Morley:2004:SAF:1018410.1018821,
 author = {Morley, David and Myers, Karen},
 title = {The SPARK Agent Framework},
 booktitle = {Proceedings of the Third International Joint Conference on Autonomous Agents and Multiagent Systems - Volume 2},
 series = {AAMAS '04},
 year = {2004},
 isbn = {1-58113-864-4},
 location = {New York, New York},
 pages = {714--721},
 numpages = {8},
 url = {http://dl.acm.org/citation.cfm?id=1018410.1018821},
 acmid = {1018821},
 publisher = {IEEE Computer Society},
 address = {Washington, DC, USA},
}

@Article{Dastani2008,
author="Dastani, Mehdi",
title="2APL: a practical agent programming language",
journal="Autonomous Agents and Multi-Agent Systems",
year="2008",
month="Jun",
day="01",
volume="16",
number="3",
pages="214--248",
issn="1573-7454",
doi="10.1007/s10458-008-9036-y",
url="https://doi.org/10.1007/s10458-008-9036-y"
}

@incollection{DBLP:books/sp/map2005/Winikoff05,
  author    = {Michael Winikoff},
  title     = {{JACK\(^{\textup{TM}}\)} Intelligent Agents: An Industrial Strength
               Platform},
  booktitle = {Multi-Agent Programming: Languages, Platforms and Applications},
  pages     = {175--193},
  year      = {2005},
  crossref  = {DBLP:books/sp/map05},
  bibsource = {dblp computer science bibliography, https://dblp.org}
}

@incollection {Hindriks:09a,
  author =	 {Hindriks, Koen V.},
  title =	 {Programming Rational Agents in {GOAL}},
  booktitle =	 {Multi-Agent Programming: Languages, Tools and Applications},
  editor =	 {El Fallah Seghrouchni, Amal and Dix, J\"urgen and
                  Dastani, Mehdi and Bordini, Rafael H.},
  publisher =	 {Springer US},
  isbn =	 {978-0-387-89299-3},
  pages =	 {119-157},
  url =		 {http://dx.doi.org/10.1007/978-0-387-89299-3\_4},
  year =	 2009
}

@article{Hindriks:1999:AP:608601.608641,
 author = {Hindriks, Koen V. and De Boer, Frank S. and Van Der Hoek, Wiebe and Ch. Meyer, John-Jules},
 title = {Agent Programming in 3APL},
 journal = {Autonomous Agents and Multi-Agent Systems},
 issue_date = {November 1999},
 volume = {2},
 number = {4},
 month = nov,
 year = {1999},
 issn = {1387-2532},
 pages = {357--401},
 numpages = {45},
 url = {https://doi.org/10.1023/A:1010084620690},
 doi = {10.1023/A:1010084620690},
 acmid = {608641},
 publisher = {Kluwer Academic Publishers},
 address = {Norwell, MA, USA},
}

@incollection{DBLP:books/sp/map2005/PokahrBL05,
  author    = {Alexander Pokahr and
               Lars Braubach and
               Winfried Lamersdorf},
  title     = {Jadex: {A} {BDI} Reasoning Engine},
  booktitle = {Multi-Agent Programming: Languages, Platforms and Applications},
  pages     = {149--174},
  year      = {2005},
  crossref  = {DBLP:books/sp/map05},
  bibsource = {dblp computer science bibliography, https://dblp.org}
}

@InProceedings{10.1007/11678823_9,
author="Braubach, Lars
and Pokahr, Alexander
and Lamersdorf, Winfried",
editor="Bordini, Rafael H.
and Dastani, Mehdi M.
and Dix, J{\"u}rgen
and El Fallah Seghrouchni, Amal",
title="Extending the Capability Concept for Flexible BDI Agent Modularization",
booktitle="Programming Multi-Agent Systems",
year="2006",
publisher="Springer Berlin Heidelberg",
address="Berlin, Heidelberg",
pages="139--155",
}

@article{Sardina2011,
	Author = {Sardina, Sebastian and Padgham, Lin},
	Day = {01},
	Journal = {Autonomous Agents and Multi-Agent Systems},
	Month = {Jul},
	Number = {1},
	Pages = {18--70},
	Title = {A BDI agent programming language with failure handling, declarative goals, and planning},
	Volume = {23},
	Year = {2011}}

@inproceedings{5285116,
	Author = {M. Dastani and B. Steunebrink},
	Booktitle = {2009 IEEE/WIC/ACM International Joint Conference on Web Intelligence and Intelligent Agent Technology},
	Doi = {10.1109/WI-IAT.2009.214},
	Month = {Sept},
	Pages = {581-584},
	Title = {Modularity in {BDI}-Based Multi-agent Programming Languages},
	Volume = {2},
	Year = {2009}}

@inproceedings{Novak:2006:MBA:1160633.1160814,
	Acmid = {1160814},
	Address = {New York, NY, USA},
	Author = {Nov\'{a}k, Peter and Dix, J\"{u}rgen},
	Booktitle = {Proceedings of the Fifth International Joint Conference on Autonomous Agents and Multiagent Systems},
	Doi = {10.1145/1160633.1160814},
	Isbn = {1-59593-303-4},
	Location = {Hakodate, Japan},
	Numpages = {7},
	Pages = {1009--1015},
	Publisher = {ACM},
	Series = {AAMAS '06},
	Title = {Modular BDI Architecture},
	Url = {http://doi.acm.org/10.1145/1160633.1160814},
	Year = {2006}}

@inproceedings{Busetta2000,
	Address = {Berlin, Heidelberg},
	Author = {Busetta, Paolo and Howden, Nicholas and R{\"o}nnquist, Ralph and Hodgson, Andrew},
	Booktitle = {Intelligent Agents VI. Agent Theories, Architectures, and Languages: 6th International Workshop, ATAL'99, Orlando, Florida, USA, July 15-17, 1999. Proceedings},
	Editor = {Jennings, Nicholas R. and Lesp{\'e}rance, Yves},
	Pages = {277--289},
	Publisher = {Springer Berlin Heidelberg},
	Title = {Structuring BDI Agents in Functional Clusters},
	Year = {2000}}

@inproceedings{Ortiz-Hernandez2016,
	Address = {Cham},
	Author = {Ortiz-Hern{\'a}ndez, Gustavo and H{\"u}bner, Jomi Fred and Bordini, Rafael H. and Guerra-Hern{\'a}ndez, Alejandro and Hoyos-Rivera, Guillermo J. and Cruz-Ram{\'\i}rez, Nicandro},
	Booktitle = {Engineering Multi-Agent Systems: 4th International Workshop, EMAS 2016, Singapore, Singapore, May 9-10, 2016, Revised, Selected, and Invited Papers},
	Editor = {Baldoni, Matteo and M{\"u}ller, J{\"o}rg P. and Nunes, Ingrid and Zalila-Wenkstern, Rym},
	Pages = {117--135},
	Publisher = {Springer International Publishing},
	Title = {A Namespace Approach for Modularity in BDI Programming Languages},
	Year = {2016}}

@inproceedings{vanRiemsdijk:2006:GMA:1160633.1160864,
	Address = {New York, NY, USA},
	Author = {van Riemsdijk, M. Birna and Dastani, Mehdi and Meyer, John-Jules Ch. and de Boer, Frank S.},
	Booktitle = {Proceedings of the Fifth International Joint Conference on Autonomous Agents and Multiagent Systems},
	Isbn = {1-59593-303-4},
	Location = {Hakodate, Japan},
	Numpages = {8},
	Pages = {1271--1278},
	Publisher = {ACM},
	Series = {AAMAS '06},
	Title = {Goal-oriented Modularity in Agent Programming},
	Year = {2006}}

@inproceedings{Hindriks2008,
	Address = {Berlin, Heidelberg},
	Author = {Hindriks, Koen},
	Booktitle = {Programming Multi-Agent Systems: 5th International Workshop, ProMAS 2007 Honolulu, HI, USA, May 15, 2007 Revised and Invited Papers},
	Editor = {Dastani, Mehdi and El Fallah Seghrouchni, Amal and Ricci, Alessandro and Winikoff, Michael},
	Pages = {156--171},
	Publisher = {Springer Berlin Heidelberg},
	Title = {Modules as Policy-Based Intentions: Modular Agent Programming in GOAL},
	Year = {2008}}

@inproceedings{Rao96,
	Author = {Anand S. Rao},
	Booktitle = {Agents Breaking Away, 7th European Workshop on Modelling Autonomous Agents in a Multi-Agent World, Eindhoven, The Netherlands, January 22-25, 1996, Proceedings},
	Editor = {Walter Van de Velde and John W. Perram},
	Pages = {42--55},
	Publisher = {Springer},
	Address = {Berlin, Heidelberg},
    Publisher = {Springer},
	Series = {Lecture Notes in Computer Science},
	Title = {AgentSpeak(L): {BDI} Agents Speak Out in a Logical Computable Language},
	Volume = {1038},
	Year = {1996}}

@incollection{jason06,
	Author = {Bordini, Rafael and H\"ubner, Jomi},
	Booktitle = {CLIMA VI},
	Editor = {Toni, Francesca and Torroni, Paolo},
	Month = mar,
	Pages = {143--164},
	Publisher = {Springer},
	Series = {LNAI},
	Title = {{BDI} Agent Programming in {AgentSpeak} Using {Jason}},
	Volume = {3900},
	Year = {2006}}

@book{bordini:07,
	Author = {Rafael H. Bordini and Jomi Fred H\"ubner and Michael Wooldrige},
	Doi = {10.1002/9780470061848},
	Publisher = {John Wiley \& Sons},
	Series = {Wiley Series in Agent Technology},
	Title = {Programming Multi-Agent Systems in AgentSpeak using \textit{Jason}},
	Url = {http://jason.sf.net/jBook},
	Year = {2007}}

@inproceedings{DBLP:conf/prima/CollierRL15,
	Author = {Rem W. Collier and Sean Edward Russell and David Lillis},
	Booktitle = {{PRIMA} 2015: Principles and Practice of Multi-Agent Systems - 18th International Conference, Bertinoro, Italy, 2015, Proceedings},
	Editor = {Qingliang Chen and Paolo Torroni and Serena Villata and Jane Yung{-}jen Hsu and Andrea Omicini},
	Pages = {351--366},
	Publisher = {Springer},
	Series = {Lecture Notes in Computer Science},
	Title = {Reflecting on Agent Programming with AgentSpeak(L)},
	Volume = {9387},
	Year = {2015}}

@inproceedings{Madden2010,
	Address = {Berlin, Heidelberg},
	Author = {Madden, Neil and Logan, Brian},
	Booktitle = {Programming Multi-Agent Systems: 7th International Workshop, ProMAS 2009, Budapest, Hungary, May 10-15, 2009. Revised Selected Papers},
	Editor = {Braubach, Lars and Briot, Jean-Pierre and Thangarajah, John},
	Pages = {237--253},
	Publisher = {Springer Berlin Heidelberg},
	Title = {Modularity and Compositionality in {Jason}},
	Year = {2010}}

@inproceedings{Nunes2014,
	Address = {Cham},
	Author = {Nunes, Ingrid},
	Booktitle = {Engineering Multi-Agent Systems: Second International Workshop, EMAS 2014, Paris, France, May 5-6, 2014, Revised Selected Papers},
	Editor = {Dalpiaz, Fabiano and Dix, J{\"u}rgen and van Riemsdijk, M. Birna},
	Pages = {58--80},
	Publisher = {Springer International Publishing},
	Title = {Improving the Design and Modularity of BDI Agents with Capability Relationships},
	Year = {2014}}

\end{document}